# RIS in Cellular Networks – Challenges and Issues

**Magnus Åström, Philipp Gentner, Omer Haliloglu, Behrooz Makki, *Senior Member, IEEE*, Ola Tageman [1]**

[1]Ericsson

Corresponding author: Behrooz Makki (e-mail: behrooz.makki@ericsson.com).

**ABSTRACT** Reconfigurable intelligent surface (RIS) has been suggested to be a key 6th generation (6G) feature and was suggested to be considered as a study-item in both 3rd Generation Partnership Project (3GPP) Releases 18 and 19. However, in both releases, it has been decided not to continue with it as a study-item, and to leave it for possible future specification. In this paper, we present the rationale for such a decision. Particularly, we demonstrate the practical issues which may affect the feasibility or usefulness of RIS in cellular networks, and present open problems to be addressed before RIS can be used in practice. Moreover, we compare the performance of RIS with network-controlled repeater, the node with the most similar characteristics to RIS and which has been standardized in 3GPP Release 18. Finally, different simulations are presented to evaluate the performance of RIS-assisted networks.

## I. INTRODUCTION

With the 5$^{th}$ generation (5G) of wireless networks and beyond, the goal is to provide everyone everywhere with high quality of service (QoS). To satisfy such requirements, wireless networks utilize high bands, e.g., millimeter wave (mmW) bands, with large bandwidths, and relies on different technologies such as beamforming and network densification. Additionally, in the future, it is expected that macro base stations (BSs) may benefit from the assistance of different types of beamforming-capable nodes such as integrated access and backhaul (IAB) nodes, network-controlled repeaters (NCRs), etc.

In Releases 16-18, the 3$^{rd}$ Generation Partnership Project (3GPP) has standardized IAB as a multi-hop decode-and-forward relaying technique in which an IAB node provides not only cellular access but also backhauling in the same hardware and/or spectrum [1]-[3]. However, IAB is more complex than a normal BS and has a large coverage area. Alternative technologies, with significantly lower complexity compared to a BS, are Release 17 radio frequency (RF) repeaters [4] and Release 18 NCRs [5] which can be used for, e.g., coverage hole removal.

As a candidate technology to assist the macro-BSs, reconfigurable intelligent surfaces (RISs), or sometimes referred to as intelligent reflecting surfaces, have received substantial attention in the academia during the last decade. In general, RISs are electromagnetically active artificial structures with beamforming capabilities that can be used to reshape the propagation environment such as to improve capacity, coverage and energy efficiency.

RISs have been studied for various functionalities ranging from wireless energy transfer [6] to sensing [7], localization [8] and communication [9]-[28], and different models of RIS, e.g., active and passive, have been proposed. Compared to passive RIS which can only adjust the phase shifts, an active RIS can also amplify the received signal by few dBs. Hardware-wise, the most common solutions proposed for RISs are based on varactors, micro-electro-mechanical systems (MEMS), positive-intrinsic-negative (PIN) diodes, and liquid crystal technologies, each with different adaptation capabilities, power consumption, etc.

From the communication point of view, the main focus on the existing works is on channel estimation, beamforming/reflection design and hardware constraints/imperfections modeling, e.g., [9]-[28]; To guarantee proper performance in RIS-assisted networks, a large number of reflecting elements is required, which increases the channel estimation overhead significantly.

Thus, a proper number of elements and low-complexity channel estimation methods have been extensively studied in the literature. Here, [9] gives a survey on channel estimation in RIS-assisted networks. Also, [10] introduces a framework leveraging RIS for energy efficiency in wireless communications, specifically using a reflect beamforming algorithm to minimize power consumption. Also, [11] demonstrates an optimized algorithm for RIS-aided communication, using joint active and passive beamforming to maximize signal strength and reliability. Zheng et al. [12] present a deep learning approach for efficient channel estimation in RIS-assisted systems, employing a neural network to predict channel states accurately. Also, [13] explores the integration of RIS with machine learning techniques, particularly a weighted sum-rate maximization algorithm to enhance system performance. Finally, [14] proposes a robust RIS design that mitigates estimation errors under practical constraints by optimizing the placement and configuration of RIS elements.

With beamforming, one of the main challenges is to determine optimal beamforming design with low complexity, while keeping the error of the cascade BS-RIS-UE (UE: user equipment) channel estimation (due to, e.g., UE mobility) below a threshold [15]. Here, for instance, [16] presents a study on joint active and passive beamforming, developing a two-stage algorithm to optimize both transmit and reflective beamforming vectors for maximized signal-to-noise ratio (SNR). Also, [17] proposes a practical RIS beamforming design leveraging alternating optimization techniques to iteratively refine phase shifts, thus enhancing overall system performance. Zhou et al. [18] introduce a robust beamforming scheme that integrates RIS with multiple-input multiple-output (MIMO) systems, focusing on improving spectral efficiency and coverage. Moreover, [19] advances the field with a robust beamforming approach, addressing hardware impairments and proposing an alternating optimization method to enhance signal quality. Then, [20] explores a hybrid beamforming scheme, combining analog and digital beamforming to improve the overall spectral efficiency in IRS-assisted systems. Finally, [21] develops an adaptive beamforming design that incorporates machine learning techniques to dynamically adjust reflection parameters, significantly boosting network throughput and reliability.

Finally, given that one of the main motivations of RIS is cost reduction (although, as we explain in the following, the network-level cost reduction of RIS-assisted networks requires further realistic evaluations), hardware imperfections may affect the reflection quality of the RISs, which have been the topic of interest in various works. Here, for instance, [22] develops channel estimation methods in RIS-assisted MIMO systems operating under different short- and long-term imperfections and [23] provides a comprehensive analysis of the impact of hardware impairments on RIS-aided systems, proposing a compensation scheme to mitigate these effects. Then, [24] examines the influence of phase noise in RIS elements and introduces a robust beamforming design to counteract these imperfections. Moreover, [25] develops a practical model incorporating hardware imperfections such as finite resolution phase shifters, demonstrating its effects on system performance. Tang et al. [26] explore the impact of non-ideal RIS elements on the achievable rate, proposing a novel optimization framework to enhance performance despite these imperfections. Also, [27] investigates the effects of hardware non-linearities in RIS-assisted systems, offering a robust design to alleviate these issues and [28] presents a study on the modeling of hardware impairments in large-scale RIS, providing critical insights and a comprehensive framework to improve network reliability and efficiency. Finally, a few testbeds have been developed recently to evaluate the feasibility of RISs in sub-6 GHz [29], 10 GHz [30] and 28 GHz [31]-[33].

During the planning phases of Releases 18 and 19 of 3GPP, the main standardization organization which develops protocols for mobile telecommunications, RIS has been suggested to be considered as a study item, i.e., to define a discussion topic which investigates if RIS-assisted communication is beneficial and/or requires standardization. However, in both releases, it has been decided not to continue with RIS as a study item, and to leave it for possible discussions in future. Consequently, many companies view RIS as a 6G feature, if it is going to be ever used practically. In this paper, we present the rationale and the discussions leading to the conclusion of not continuing with RIS as a study item. Such discussions provide guidelines for the researchers on how to improve the practicality of RISs, such that they can be practically used in cellular networks. Particularly, we investigate the practical challenges of RIS-assisted networks in terms of signal reflection properties, interference management, signaling overhead, cost, deployment, network integration as well as regulatory aspects. We study propagation modeling, the effect of hardware imperfections as well as the intermediate field effects in RIS-assisted networks. Moreover, we compare the performance of the RIS with NCR, the node type with most similar characteristics to RIS which has been standardized in 3GPP Release 18. Finally, we present different simulation evaluations to evaluate the performance of RIS-assisted networks, in comparison with alternative technologies. As we show, while RIS looks interesting on paper, there are various practical issues to be addressed before it can be used in practical cellular networks.

## II. NCR as an Already-Standardized Relay

3GPP has defined and specified the requirements/capabilities for NCR in Release 18. In simple words, NCRs are normal amplify-and-forward repeaters with beamforming capability which can receive and process control information from the network. An NCR is deployed and under the control of a mobile network operator (MNO) and, for all management purposes, is logically part of its controlling gNodeB (gNB). Intuitively, an NCR has no channel and signal awareness, and always follows the gNB commands to forward the signals back-and-forth between the gNB and the UE in downlink (DL) and unlink (UL) with proper beamforming. In this way, an NCR can be thought of as a network-controlled *beam-bender* relative to its controlling gNB.

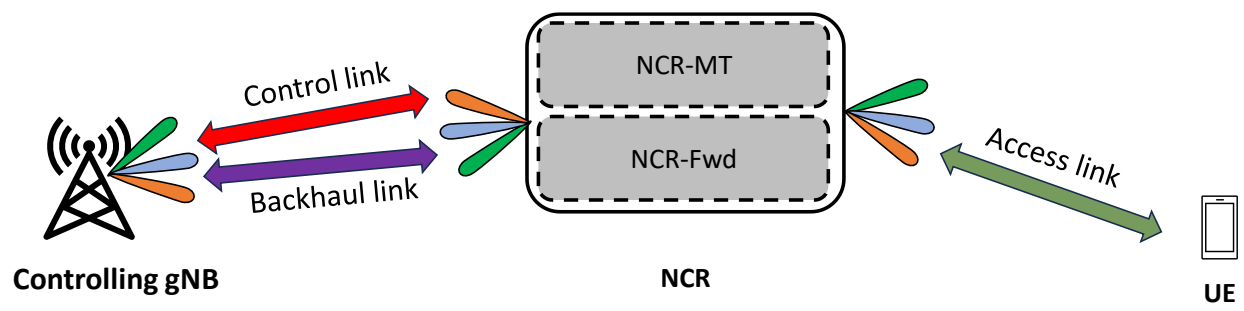

**Figure 1. The schematic of an NCR defined in 3GPP Release 18.**

NCRs are of interest in different indoor and outdoor networks as well as for indoor-to-outdoor (I2O) or outdoor-to-indoor (O2I) communications. In 3GPP Release 18, the NCR has been specified for single-hop communication in stationary deployments considering both FR1 (sub-6GHz) and FR2 (mmW) bands [5]. The NCR is transparent to the UE, i.e., the UE does not notice the presence of the NCR and consequently there has been no UE specification impact from the NCR specification work.

Figure 1 shows the structure of the NCR, as defined by 3GPP Release 18 [5]. An NCR consists of two modules:

- NCR-mobile termination (NCR-MT) is a module to communicate with the controlling gNB to exchange side control information via the control link.
- NCR-forward (NCR-Fwd) is a module to amplify-and-forward UL or DL signals between the gNB and the UE.

Backhaul and access links refer to the links between a gNB and an NCR-Fwd and between an NCR-Fwd and a UE, respectively. The information exchange between the gNB and the NCR enables efficient amplify-and-forwarding in the NCR. Particularly, the operation of the NCR-Fwd is controlled based on the control information received by the NCR-MT from the controlling gNB. Such side control information mainly includes information about proper beamforming, time division duplex (TDD) operation and ON/OFF configuration of the NCR. In other words, the side control information configures the NCR to use specific beams at specific time resources to forward the signals in either DL or UL, while the NCR has no knowledge about the forwarded signals.

There are similarities and differences between NCRs and RISs. As the main similarity, both nodes forward the signals in predetermined directions without decoding them. The received signal is immediately forwarded, i.e., there is no half-duplex constraint. Different from an NCR though, a RIS does not amplify the signal (or, in the cases with an active RIS, RISs only amplify the signal by few dBs as opposed to NCRs with up to 90-100 dB amplification gain). As a result, RISs do not amplify-and-forward the noise/interference, which is an advantage of RISs over NCRs. Finally, although there is yet no standardized structure for RISs, it is expected to follow similar schematic as in Fig. 1 where a module will be used for information exchange with the controlling gNB enabling proper RIS configuration. In Sections IV and V, we present more detailed qualitative and quantitative comparisons between NCRs and RISs, respectively.

## III. Challenges with RIS in Cellular Networks

RIS has gained significant academic interest over the last couple of years and is promoted by some as a candidate component in the future 6G cellular networks.

For a novel feature like the RIS to be interesting for a commercial network, it must provide clear benefits without adding too high costs such that the net is clearly positive. In this paper, we assume a network that does not deviate too much from the present New Radio (NR) specification, an assumption that we think will hold true also for the 6G networks. We further assume a RIS, similar to the NCR, to be an extension of a BS and hence a part of the cellular network, i.e., a network-controlled RIS (NC-RIS). In this way, the system model of the RIS-assisted network considered in this paper can be represented as in Fig. 2.

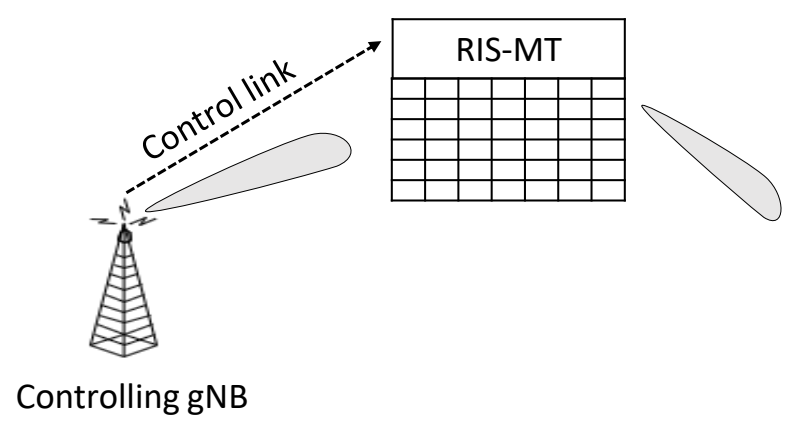

**Figure 2. An example of the system model considered in the paper.**

### A. Target Spectrum

RIS is able to provide coverage within a cell's coverage range where, for some reason, the BS is not able to provide coverage or sufficiently good throughput. Scenarios where this may be the case is O2I or non-line-of-sight (nLoS) outdoor. We note that blockers are highly frequency dependent where FR1 spectrum typically have good O2I penetration whereas even foliage may be sufficient to create an nLoS link in FR2. That makes FR2 (or FR3 in 7-15 GHz bands) more suitable spectrum for RIS compared to FR1, since FR2 inherently exhibits more spotty coverage and less channel richness due to the mmW signal properties whereas FR1 spectrum naturally exhibits robust coverage and multi-path environments.

Regardless of spectrum, a (passive) RIS is, however, typically unable to extend coverage beyond the cell edge since it does not amplify a signal but only redirects it. An exception is possibly that the backhaul link between the BS and the RIS could be better than what would be the case for a direct link BS to UE, in which case two longer line-of-sight (LoS) BS-RIS and RIS-UE links would be preferrable to a single nLoS BS-UE link.

### B. Signal Reflection Properties

Contrary to the NCR, a RIS is acting as a (passive) reflector. The overall free space pathloss for a sufficiently large, flat and ideal reflector is proportional to the square of the sum of the distances of the individual links, BS-RIS and RIS-UE and on par with the direct LoS path. However, in practice, the reflector with its aperture size is limited as presented in Sec. V, where a more realistic RIS model shows that the pathloss (in dB) from a reflected LoS path will nearly double compared to the single LoS path for the same total distance. However, by introducing a third dimension, depth, focusing the signal energy at a certain distance, the RIS may hypothetically focus a signal on the UE location assuming the RIS's near-field properties applies. Such a focusing, however, will cause problems with mobility, limiting the practical use cases (See Section IV for more details). Additionally, it would add another degree of freedom (depth) to the already high RIS control overhead. However, for both the near-field and far-field scenarios, the RIS is a far from perfect reflector. For the far-field case, the reflection is at best on par with a metal plane with equal size and some types of RIS provide lots of side lobes that will act as interference in their respective directions. Tellingly, much of the literature compares RIS reflection performance to that without the RIS, disregarding the relevant question about how much energy is constructively reflected by the RIS in the desired reflection direction and how much is reflected elsewhere as interference.

In order to reduce the impact from limited reflection performance and taking advantage of their low cost, RISs are expected to be large, at the expense of high total cost of ownership (TCO), overhead, etc.

### C. Interference Properties

Being a surface, a RIS is, in effect, a spatial linear transform operating on any incidence wave, resulting in arbitrary reflection waves. This makes it prone to propagate interference instead of attenuating or cancelling it. Naturally, a beam targeting the RIS is expected to be (one of) the strongest waves but other waves will dynamically cause reflections that are not expected or accounted for in the network, thereby reducing overall network performance.

In addition to the unrestricted spatial properties of a RIS, since it does not contain any spectral filtering functionality, it has equally unrestricted spectral properties. This implies that the unintended dynamic reflections resulting from a RIS may interfere in a much wider spectrum than what is usually the case with cellular communications. Particularly, a RIS will not only act as a reflector in the spectrum for which it is intended but also dynamically and randomly affect a wide spectrum around it. In our understanding, that highly unattractive property is so far not properly addressed by RIS proponents, e.g., [34].

A less critical but still not negligible property, in particular of resistance and capacity (RC)-based RIS implementations, is non-linear effects appearing from saturation in the circuitry if the incidence wave is too strong. As a result, adjacent carriers would be directly affected in a way that up until now has been prohibited.

A substantial part of developing a new standard is to ascertain that network nodes can co-exist both within a network and among nodes belonging to different networks, i.e., that interference is restricted. In 3GPP, this is typically RAN4's responsibility. Outside of 3GPP, regulatory bodies may add further restrictions. Up until now, large efforts have been made both in standardization and implementation to limit both co-channel, adjacent channel and out-of-band interference. RIS risks being a departure from that principle.

### D. RIS Signaling Overhead

RIS will involve two kinds of signaling overhead. First, the complexity added to reference signaling in order to support RIS, second, the complexity in configuring and controlling the RIS itself.

While the present networks can be approximated to have three degrees of freedom for determining the Tx and Rx beams at the BS and UE, respectively – vertically and horizontally at the BS and linearly at the UE – a RIS network may significantly increase that. A cell may include multiple RISs, adding another degree of freedom and as discussed elsewhere in this paper, each RIS may in turn require narrower beams, i.e., requiring multiples of reference signals per degree of freedom for the BS. All in all, this leads to substantially higher configuration overhead.

As presented above, in order to be commercially attractive, a RIS needs to cover a substantial area which, in turn, requires a sizable number of elements in the RIS to achieve sufficient reflection power in the desired direction. Thousands of elements and orders of magnitude more are not uncommon in the literature, e.g., [35]. Even if the degrees of freedom for control of such elements are severely restricted in that all combinations will not result in a coherent reflection wave, and even certain wave directions may be redundant, the signaling overhead of such a node may still be substantial for a number of reasons.

In our view, there are a few cases that particularly stand out as problematic for RIS in a cellular network:

- **Cell-specific signals**, in limited supply, a substantial amount of which would need to be allocated to RIS,
- **Mobility tracking signals** and radio resource management (RRM) complexity would increase due to the narrower RIS beams, and
- **RIS control signaling** increases from the necessarily increased resolution of configuration parameters.

The implication with respect to overhead this will have for RIS is that:

the larger the required coverage area Þ
the larger the RIS Þ
the narrower the beams Þ
the larger the necessary overhead.

The above relations will, in practice, set a limit on the number of RISs per BS and the size of each RIS. The same can of course be said for the NCR, however, the operating point between overhead and coverage differs in NCR's favor in that the NCR requires less overhead for the same provided coverage (see Section IV for more discussions).

Putting into the present 5G NR context, the specification limits the number of synchronization signal blocks (SSBs) to 64 in a cell. Assuming Fast Fourier Transform (FFT)-based reflection beams in the horizontal dimension, the RIS would be limited to 64 columns – a not very impressive RIS. Additionally, all the BS's SSB resources would be consumed by this single RIS. There are of course some remedies to this, like configuring multiple primary carriers, and all RISs may not be configured to reflect SSBs, but this exemplifies one inefficiency with RIS that needs to be addressed before RIS can be flexibly and efficiently deployed in networks.

### E. Deployment Aspects

When RIS is discussed as a 6G feature, cost is often brought up as an advantage, claiming RIS is a low-cost node type. That may be true if only the RIS production cost is considered, but less so if the TCO is accounted for. The differences between the two are, e.g., planning, permitting and deployment costs, cost for power and site rent. As a relatively low performing node type (compared to a BS or a repeater), the RIS is not allowed to incur high costs during any stage in its life cycle.

Although a RIS may be cheaper than an NCR, to cover an area, there may be a need for fewer NCRs, compared to RISs. Moreover, even if the RIS itself is cheaper than an NCR, a large part of the TCO is related to installation, site rental, etc. which is indifferent of the node type (see Fig. 3). In fact, the larger size of the RIS may even result in, e.g., higher site rental/installation cost, compared to, e.g., an NCR. Thus, to have a realist view, there is a need for deep *network-level cost-performance* analysis in the area considered.

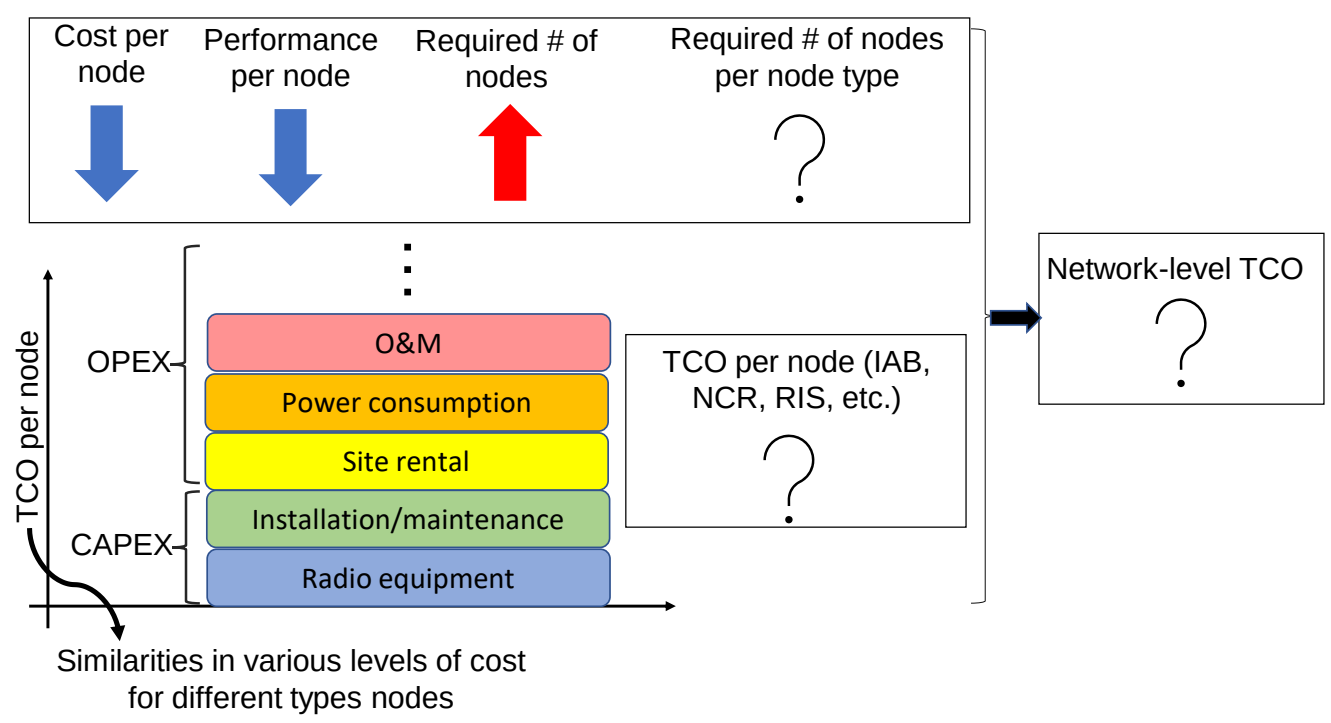

**Figure 3. Network-level TCO analysis for different types of nodes. The figure is used as an illustrative example, and in reality, the relations between different levels of cost does not follow the figure. Ericsson has no views to be presented externally about the per-node or network-level TCO of different types of nodes.**

One consideration that must be accounted for when introducing a novel node type is that it is sufficiently versatile to justify not only development of the RIS itself but also implementation and integration in the network and possibly also UEs. For RIS, this means, e.g., that deployments need to be sufficiently versatile such that the plane on which the RIS is deployed, e.g., a building wall, allows the RIS to support the intended coverage area. A related matter is the immediate area around the RIS. For a RIS to operate efficiently, it is advantageous for it to establish LoS links both towards the BS and the UE. Both the deployment plane and surrounding area will constrain the number of locations where it is practically feasible to deploy a RIS. In a dense urban scenario, this limitation is likely not insignificant and hence a clear disadvantage for RIS.

One kind of RIS that has been proposed is the translucent RIS for O2I deployments, possibly integrated in windows [31]. Although technically feasible, it is worth noting that not all kinds of businesses replace their output after 10 to 25 years, which is the typical life span of a network node and generation of cellular networks, respectively. Some sooner, others later. Building construction is one business that operates on much longer time constants than the telecom industry – a house can stand for hundreds of years. Assuming the RIS to be integrated into buildings will bring extra costs in that not only does the RIS need to cover its own cost but also the cost for replacing the construction material much earlier than what would otherwise be the case. As an example, a window, with a typical life span of 50 years, may need replacement after only 10 years when the RIS has reached its end of life.

### F. Network Integration

In order for the RIS to become an integrated part in cellular networks, it must comply with the most arduous network requirements. Although LTE and NR share modulation technique and thereby have similar link performance, latency is substantially reduced in NR. From a network perspective, it is clearly disadvantageous to integrate a node type like the RIS with properties that are substantially worse than what is expected in today's or tomorrow's networks. Hence, e.g., the current LCD-based RIS with switching times in the range of milliseconds may be problematic to integrate in networks operating on microsecond time constants.

### G. Regulatory aspects

A matter that has not gained much attention related to RIS is the regulatory aspects of RIS. The above presented issues with RIS, both as a potential interferer outside its intended carrier (and likely directly affecting competing MNOs) as well as being a node type that can manipulate electromagnetic waves to dynamically focus the energy locally, is all but certain to gain the interest from regulatory bodies even if the RIS itself is a passive node. Another aspect is who is responsible for certification of RIS in case of intermodulation interference. The MNO, the RIS vendor or the BS vendor? Considering the novelty of the RIS, regulatory aspects are likely to extend over many years before any decision is made.

## IV. Comparison between NCR and RIS

From a practical point of view, NCR has properties which, compared to RIS, make NCR more attractive in cellular

networks:

- RIS has no spatial selectivity whereas NCR does. That is, RIS will reflect signals from any direction whereas NCR's beamforming gain efficiently provides spatial filtering. This is a first disadvantage for interference management.
- RIS has no spectral selectivity whereas NCR does. That is, RIS will reflect adjacent carriers or even out-of-band signals equally well as inband signals. This is a second disadvantage for interference management.
- To guarantee the same performance as in NCR, RIS requires a significantly larger number of reflecting elements, e.g., [36]. As a result, RIS is (potentially) large, which makes its deployment more challenging, compared to an NCR.
- Due to the large number of elements in the RIS, the signaling overhead in RIS is (potentially, much) larger than NCR.
- The efficiency of the RIS is considerably affected by the double path loss effect in the gNB-RIS-UE link as well as the near necessity of (nearly) LoS links in both link segments. With an NCR, however, the double path loss effect is partly compensated by the high amplification gain of the NCR.
- Moreover, an RIS has only one phase shift matrix, while an NCR performs separated beamforming at gNB- and UE-sides. This gives the NCR a better capability for, e.g., interference management.
- Deployment for RIS may be further complicated in that both links need to be LoS in the vicinity of the RIS location. In an NCR, on the other hand, the backhaul link and access link can be separated by several meters, allowing for more versatile deployments.

For quantitative comparisons between the NCRs and RISs, see Section V.

## V. Simulation Results

In this section, several simulations results are presented including discussions on propagation modeling, the effect of hardware impairments, intermediate-field effects as well as comparison of RIS with alternative technologies.

### A. Basic Propagation Modeling

When a RIS is located at large enough distance from both the BS and the UE, it is relevant to make a far-field approximation and treat the RIS as a scatterer from incoming plane waves to outgoing plane waves, as illustrated in Fig. 4. To estimate the maximum achievable end-to-end gain between BS and UE, the RIS can be modeled as a dual polarization phased array antenna with phase shifters connected to its ports – shorted to reflect signals with a tunable phase shift. It is not necessary to separate out the phase shifting from the antenna in this way, but it will allow for a more intuitive interpretation. Dissipative loss and reflection due to impedance mismatch is ignored, and elements are assumed to be placed at half-lambda spacing to avoid grating lobes. More specifically, the phase shifters do not create any reflection other than the intended total reflection at the shorted end, and the array antenna provides perfect active match for plane waves. "Active" relates to the condition that all ports are excited simultaneously in accordance with transmission or reception of a planewave in a certain direction. Mutual coupling between elements plays an important role and must be considered during optimization of the antenna element towards perfect active match [37]. Phase shifters are assumed to have infinite resolution and to produce a constant phase gradient that matches the incoming and outgoing propagation directions. Under these conditions: reception, phase shifting, and transmission will occur in a single sequence, such that the RIS can be accurately modeled as two separate array antennas with phase shifters in between (with two times the original phase shift). Amplitudes are assumed to be uniform both at reception and transmission, because of the plane-wave assumption and absence of reflections and insertion loss.

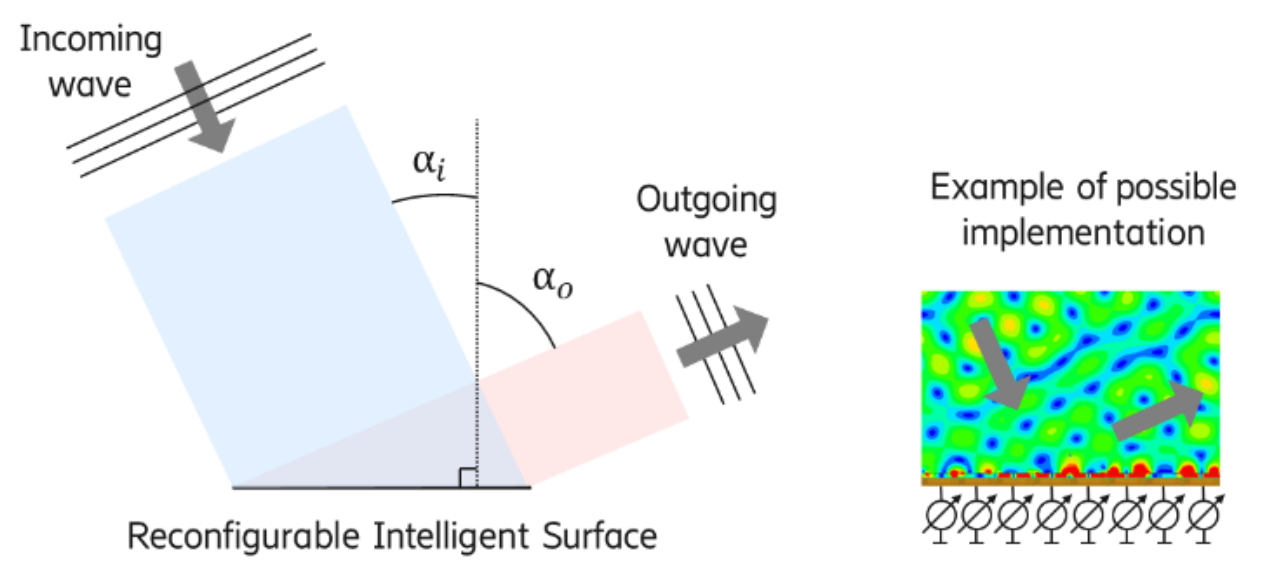

**Figure 4. Illustration of wave-propagation near the surface of a portion of a RIS for a far-field case. For large angles the projected area is considerably smaller than the actual area.**

The gain of such phased array antennas is readily estimated, from total area, A, the wavelength, λ, and the steering angle relative to the surface normal ($\alpha_i$ and $\alpha_o$), under the assumption that they are large compared to the wavelength [37]. For non-zero steering angle, it is the projected area rather than the actual surface area that matters, as illustrated in Fig 4. It should be noted that the gain cannot be increased by further sub-division into smaller elements (for fixed total area).

To further simplify the discussion, the receiving and transmitting phased arrays can, for given steering angles, be modeled as two interconnected fixed beam antennas with gain corresponding to the projected areas mentioned in the previous paragraph. This model is shown in Fig 5. Friis transmission equation can then be applied two times to find the total end-to-end gain between the BS and the UE.

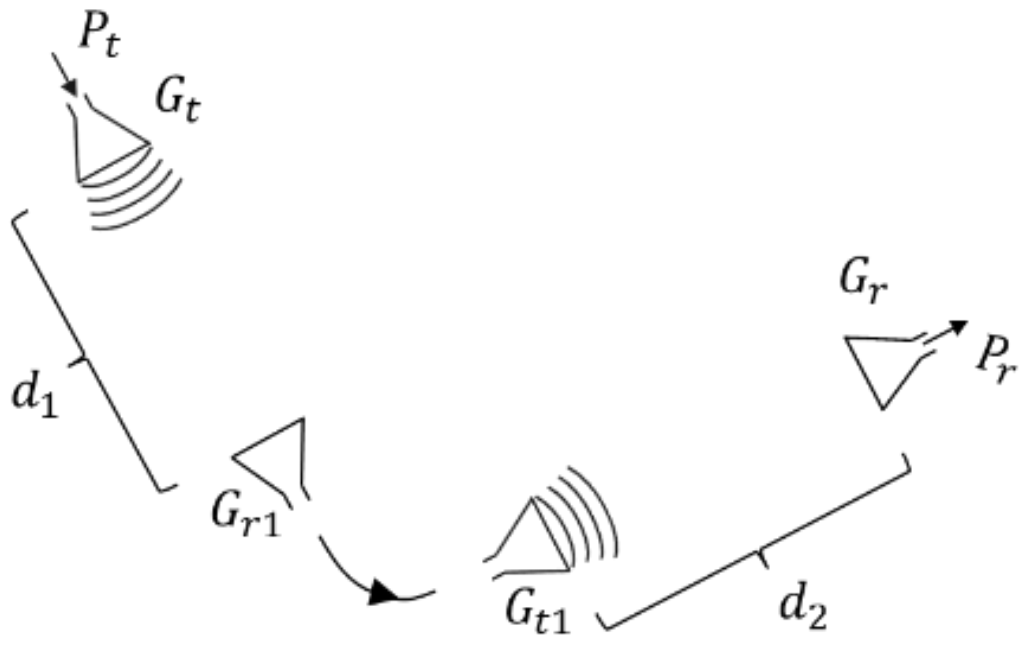

**Figure 5. Simplified propagation model useful for estimating gain, end-to-end.**

The expression for total end-to-end gain can be divided into three contributions: one that has to do with the free space path loss, one that has to do with the RIS antennas ($G_{r1}$ and $G_{t1}$ in Fig, 5), and one that has to do with the BS- and UE-antenna ($G_t$ and $G_r$ in Fig. 5). In the following, the two first gain contributions will be denoted by "Free space path loss" and "RIS-gain", respectively, while the third contribution will be disregarded since it simply adds a constant gain that is independent of the RIS configuration (at far field distance).

Four different deployment cases are defined in Table I, along with two reference cases without RIS, for the purpose of making it clear how much better or worse the end-to-end gain becomes when bouncing off a RIS compared to the corresponding LoS case, given the same total propagation distance (d1+d2). One might argue that it is unfair to compare with a direct LoS hop, since a RIS would not be installed unless the LoS-path is obstructed, but the LoS case will at least provide an absolute reference with a clear and simple definition.

Table I.
Deployment cases to be studied.

| Case | Freq. | Distance | Size |
|---|---|---|---|
| 1S | 30 GHz<br>$\lambda$=1cm | 100+100 m<br>d1=10000 $\lambda$, d2=10000 $\lambda$ | 1000 elements<br>250$\lambda^2$ (0.16x0.16 m$^2$) |
| 1L | 30 GHz<br>$\lambda$=1cm | 100+100 m<br>d1=10000 $\lambda$, d2=10000 $\lambda$ | 10000 elements<br>2500$\lambda^2$, (0.5x0.5 m$^2$) |
| 2L | 30 GHz<br>$\lambda$=1cm | 180+20 m<br>d1=18000 $\lambda$, d2=2000 $\lambda$ | 10000 elements<br>2500$\lambda^2$, (0.5x0.5 m$^2$) |
| 3S | 3 GHz<br>$\lambda$=10cm | 180+20 m<br>d1=1800 $\lambda$, d2=200 $\lambda$ | 1000 elements<br>2500$\lambda^2$, (1.6x1.6 m$^2$) |
| 4 5 | 30(3) GHz | 200 m LoS | No RIS |

In Fig. 6, the free space path loss is shown as a function of distance (normalized to the wavelength), both with the RIS placed in the middle (with d1 equal to d2 in Fig. 5) and with the RIS placed closer to one end (with d1 nine times larger than d2). As a reference, the gain for direct LoS hop without RIS is included in Fig. 6. Note that the total distance is kept the same in all cases. The deployment cases from Table I are marked with circles in the figure.

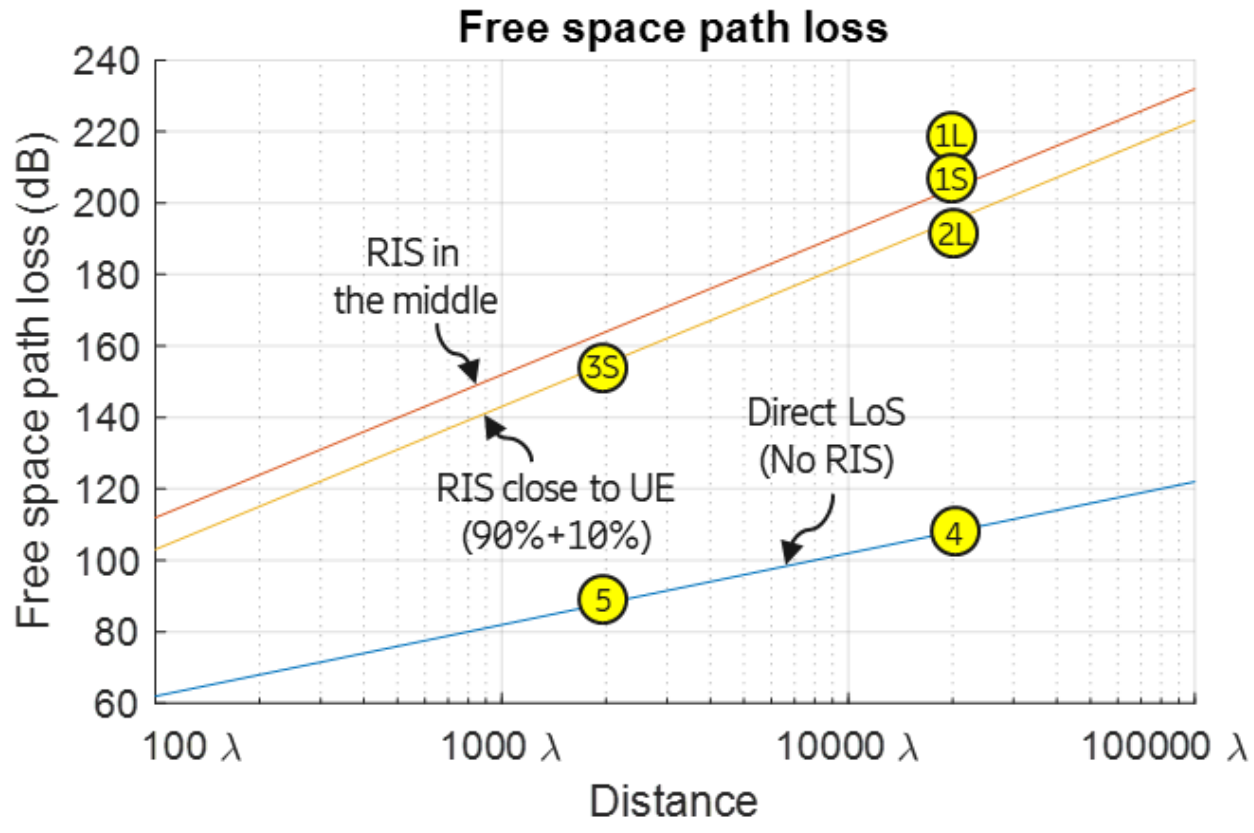

**Figure 6. Free space path loss for a single hop, two step hop and two step hop with the RIS moved closer to the UE.**

As seen from Fig. 6, a two-step hop will almost double the free space path loss (in dB) compared to a single hop (direct LoS across the same total distance). To be exact it is 12 dB less than doubled, when the RIS is placed in the middle (which is related to the halving of each distance in Friis transmission equation). In principle, if the RIS is large enough it can compensate for the increased free space path loss. Unfortunately, increased size will lead to increase cost and complexity, and it will make mobility worse due to the shrinking beam width. This trade off will explored in following.

Figure 7 shows the RIS gain ($G_{r1}$ and $G_{t1}$ in Fig. 5 combined) as a function of area (normalized to λ squared). To be precise one should, instead of the actual surface area, use the projected areas related to $\alpha_i$ and $\alpha_o$ when reading the figure and allow the two to be different. With these details ignored the RIS-gain is overestimated by 3 dB at most, provided that the incident and outgoing angles never exceed 45 degrees. Three different traces are shown in Fig. 7 to illustrate how the distance comes into play. The straight line is valid when the UE is in the far-field region of the RIS. The bent traces illustrate what occurs if the distance to the UE is reduced as indicated by the labels in the figure, in case the RIS is unable to create a curved phase front. This gain drop arises in the summation of complex valued field contributions from all elements across the RIS when elements near the edge acquire significantly more phase shift than those in the center. If, on the other hand, the RIS is capable of creating a curved phase front, it can eliminate the gain drop even at short distance and recover the far-field RIS-gain (straight line). It should be noted that a simple flat metallic plate with optimal alignment in both elevation and azimuth provides optimal performance at far-field distance, and any attempt to focus the beam will lead to reduced gain. The deployment cases from Table I are marked in Fig. 7, except for Cases 4 and 5 since there is no RIS in those cases.

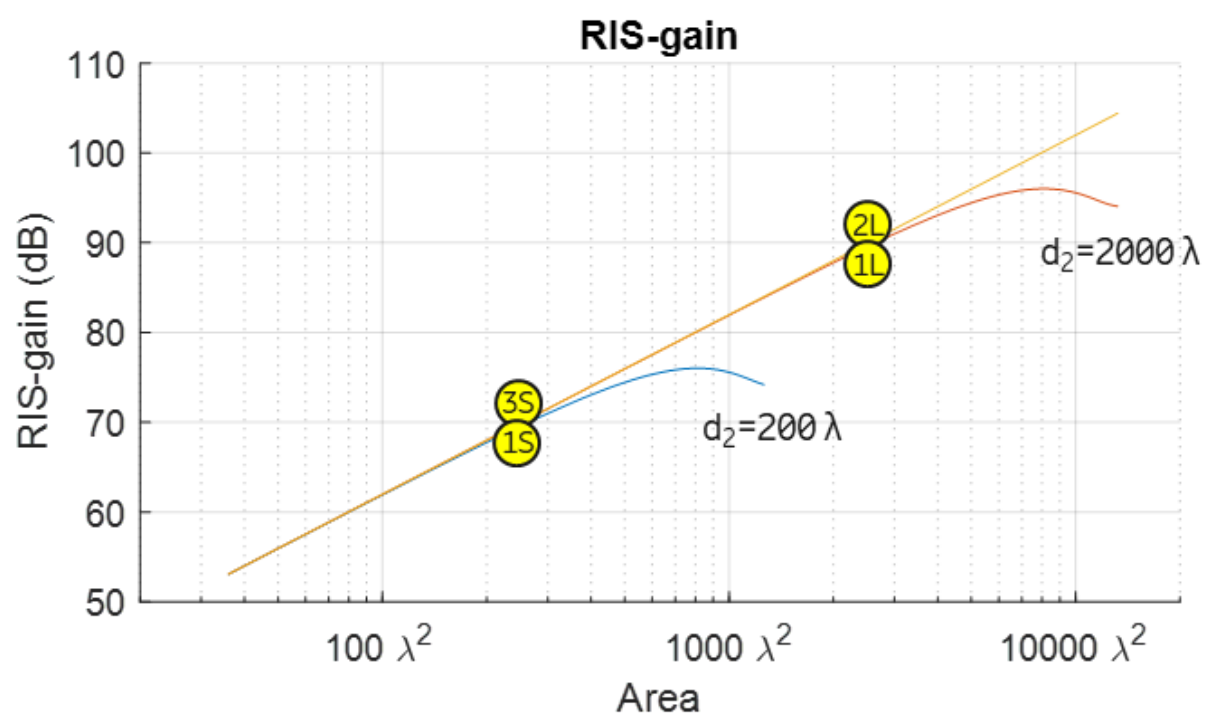

**Figure 7. RIS-gain as a function of area, ignoring loss, mismatch and gain drop due to the projection effect.**

A disadvantage of making the area larger is that the half-power beamwidth will get smaller. A small beamwidth will at short distance severely restrict UE movement, sideways and vertically. From the half-power beamwidth (angle) in boresight for a large circular array with uniform excitation [38], it is straightforward to calculate a corresponding half-power beam diameter (length) at a certain distance from the RIS. This beam diameter is shown as a function of RIS-area for three different distances in Fig. 8. Off-boresight one should, to be more exact, use the projected area when reading the figure. This will at most increase the half power beam diameter by 40 %, provided that the outgoing angle never exceeds 45 degrees.

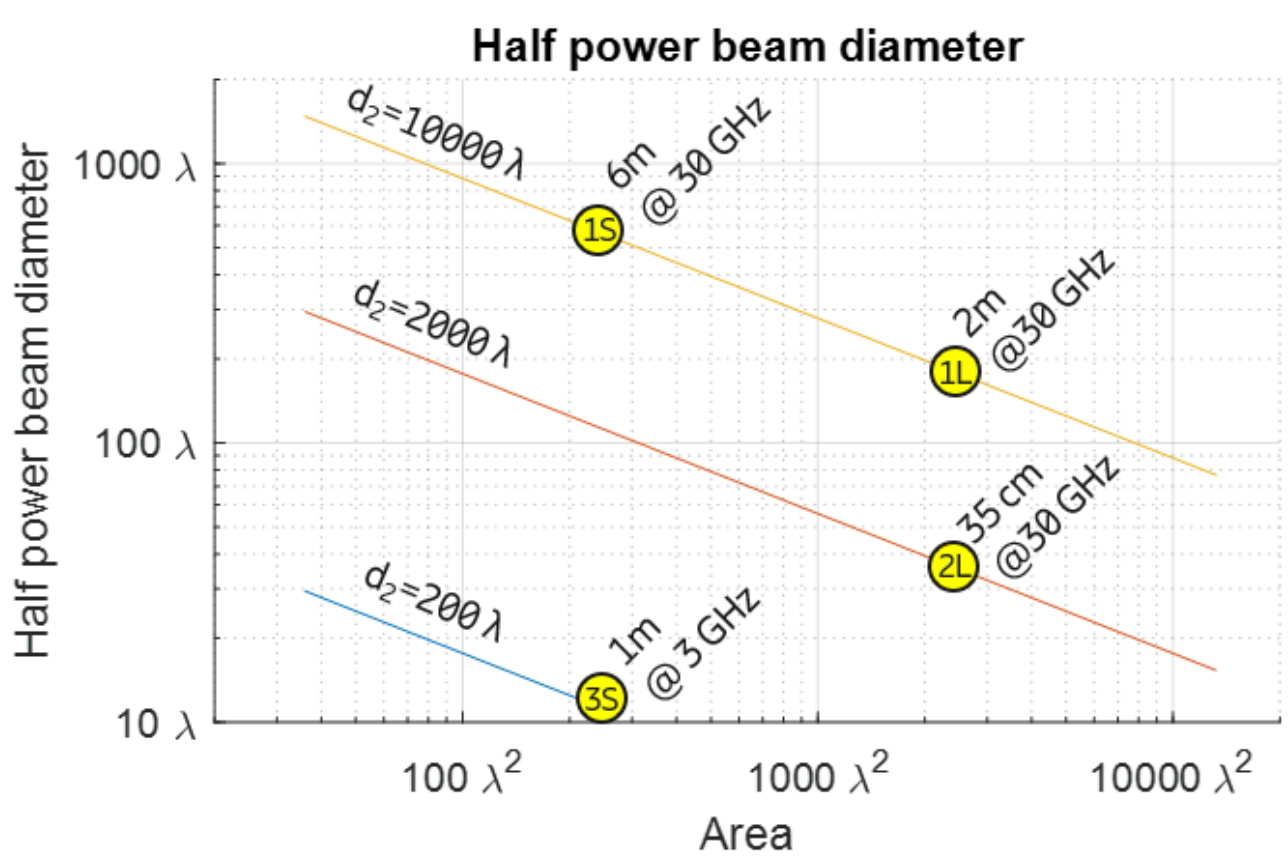

**Figure 8. Half-power beam diameter at a certain distance from the RIS as a function of the area of the RIS.**

Table II provides a comparison between the different deployment cases. Note that the table assumes ideal conditions, with loss ignored, and only lists boresight results. The “Gain relative to LoS-case” is found by subtracting the free-space path loss from the RIS-gain and relating it to the free space path loss for the LoS-case. Case 1S, which considers a 1000-element, 30 GHz RIS located halfway between the BS and the UE, gives 26 dB lower end-to-end gain than the reference case (LoS without RIS). Case 1L has 10000 times more elements which improves the gain to 6 dB below the reference case. While this would be a useful gain level, there is a significant penalty in terms of complexity and mobility to have 10000 elements and a beam diameter less than 2 m. In an attempt to further increase gain, one could put up more RIS-panels such that there is always one to find within 20 meters from the user. This is the assumption in Case 2L. In this case, the end-to-end gain exceeds the gain in the reference case by 3 dB. A clear drawback of this is that the beam diameter is decreased down to 35 cm, which further reduces the support for mobility. In addition, such a high density of RIS-panels does not appear realistic across large areas.

Another way to improve the gain is to turn to lower frequencies, such as in Case 3S (3GHz). At this frequency, the size of the RIS-panel gets strikingly large, 1.6 m x 1.6 m, already for 1000 elements. With the RIS located 20 meters away from the UE the gain exceeds that of the reference case by 3 dB. However, having such large RIS-panels within 20 meters from all UE locations of interest appears unrealistic. Furthermore, the beam diameter is 1 meter, which again will give poor mobility. In addition, 3GHz is not a particularly interesting case, since diffraction is more effective at this low frequency. Considering the above discussion, we fail to identify a sweet spot where RIS makes good sense, but we tend to consider Case 1L as a kind of reference case that gives the least unattractive trade-off between complexity, gain, and mobility.

Table II.
Final comparison between studied cases, which exemplifies the trade-off between size (complexity, cost), gain and half power beam diameter (mobility).

| Case | Freq. | Distance | Size | Gain relative to LoS -case | Beam diameter |
|---|---|---|---|---|---|
| 1S | 30 GHz | 100+100 m | 1000 elements 0.16x0.16 m² | -26 dB | 5.5 m |
| 1L | 30 GHz | 100+100 m | 10000 elements 0.5x0.5 m² | -6 dB | 1.8 m |
| 2L | 30 GHz | 180+20 m | 10000 elements 0.5x0.5 m² | +3 dB | 35 cm |
| 3S | 3 GHz | 180+20 m | 1000 elements 1.6x1.6 m² | +3 dB | 1.1 m |
| 4 5 | 30 (3) GHz | 200 m LoS | No RIS | 0 (reference) | 11 (35) m for 35 (30) dBi BS |

## B. Hardware Imperfections

In practical implementations, hardware imperfections will lead to performance degradation [22]. The need for large size (1000-10000 elements) and low cost will likely force the designer to stay with few and thin layers of inexpensive materials for the RIS elements and signal routing, and rely on simple manufacturing methods, and low-cost technologies for phase tuning, control signaling and data processing – all with a low power consumption. This will compromise bandwidth, impedance match, insertion loss, phase errors, response time, allowed complexity, etc.

The RIS elements will not be perfectly matched for plane-waves at all angles and frequencies, which will lead to reflections that will disturb the excitation amplitudes and phases, which in turn will increase emissions in unwanted directions. Of particular concern is the uncontrolled emission in the mirror direction, which can potentially interfere destructively with the main beam, if that is steered in the same direction. Limited phase shifter resolution will break the assumption of constant phase gradient, which will lead to a gain drop in the main beam and increased emissions in unwanted directions, in analogy to grating lobes for large sub-arrays. Side lobe levels will rise if there are significant phase errors due to variations in manufacturing, which may require sophisticated calibration approaches. Insertion loss in the RIS elements and tuning arrangement will reduce the RIS-gain, and this can be significant at 30 GHz. Slow adaptation to changes in UE-position, or the propagation environment will limit mobility. The response time is limited by the response time of the tunable device technology, the time it takes to propagate out changes in bias to them, and the time it takes for algorithms to estimate what the best set of bias settings is.

Possible technologies for implementing a RIS include PIN and varactor diodes, RF-MEMS, photo conductive and Ferro-electric switches, and liquid crystal technology [39]. With the key performance indicators (KPIs) of operating frequency and voltage, power consumption, switching speed and cost, one will face with tradeoffs for the whole system. Also, not all technologies are suitable for FR1, FR2 or FR3 bands.

### *C. Intermediate Field Effects*

Intermediate field effects will not play a significant role in any of the deployment cases listed in Table 1. The RIS gain in all cases stays close to the straight line corresponding to the far field limit in Fig 7. In more extreme cases, with even shorter distance and/or larger RISs, one should take intermediate field effects into account. The gain reduction that arises can, if needed, be eliminated by creating a curved phase front that creates focusing at a finite distance from the RIS. This can be thought of as an adding a third dimension, distance (or depth), to the beam-space to select from, which is likely to further increase the complexity of search algorithms and thus reduce mobility or increase overhead. Furthermore, it would in principle be possible to have independent channels to different UEs in one and the same direction, if they are at different distances, and one could even exploit LoS MIMO to a single user, if its antenna is large enough. However, the benefits would be limited since when dividing the signal into several paths on the UE side of the RIS, one would have to divide the total signal power and capacity that the link between the BS and UE provides.

If a RIS, or in general any reflecting essentially flat surface (e.g., building wall), is comparable in size to the beam diameter or larger, a far-field approximation is not justified. Both phase- and amplitude variations must be considered in this case, and a curved phase front is required if gain is to be optimized. Ultimately when a flat reflecting surface is much larger than the beam diameter the end-to-end gain including the reflection will be the same as for a single hop across the same total distance (d1+d2), which is non-optimal (since the surface is flat).

### *D. Comparisons with Alternative Technologies*

A blind spot coverage scenario has been investigated for RIS and other competitive technologies, i.e., basic metal reflector which does not change the amplitude but randomly changes the phase of the incoming signal and an NCR, i.e., a full-duplex relay with amplify-and-forward (AF). A single BS with $M$ antennas, single NCR/RIS with $N$ antennas/elements are deployed in a rectangular area of 160 x 100 meters, and a single-antenna UE sweeping the UE location with 5 meters inside the area as shown in Fig. 9. It is assumed that direct BS-UE path is blocked due to blocker, and the UE is served via NCR or RIS. It is assumed that BS-NCR/RIS link is LOS and BS-UE link is nLoS. Perfect channel knowledge and no hardware impairments are assumed. The effects of quantized phase shifts at RIS, self-interference at NCR, external interference for RIS and NCR have been considered. The simulation parameters are given in Table 3.[1] The models for different types of nodes considered in Figs. 9-11 are as follows.

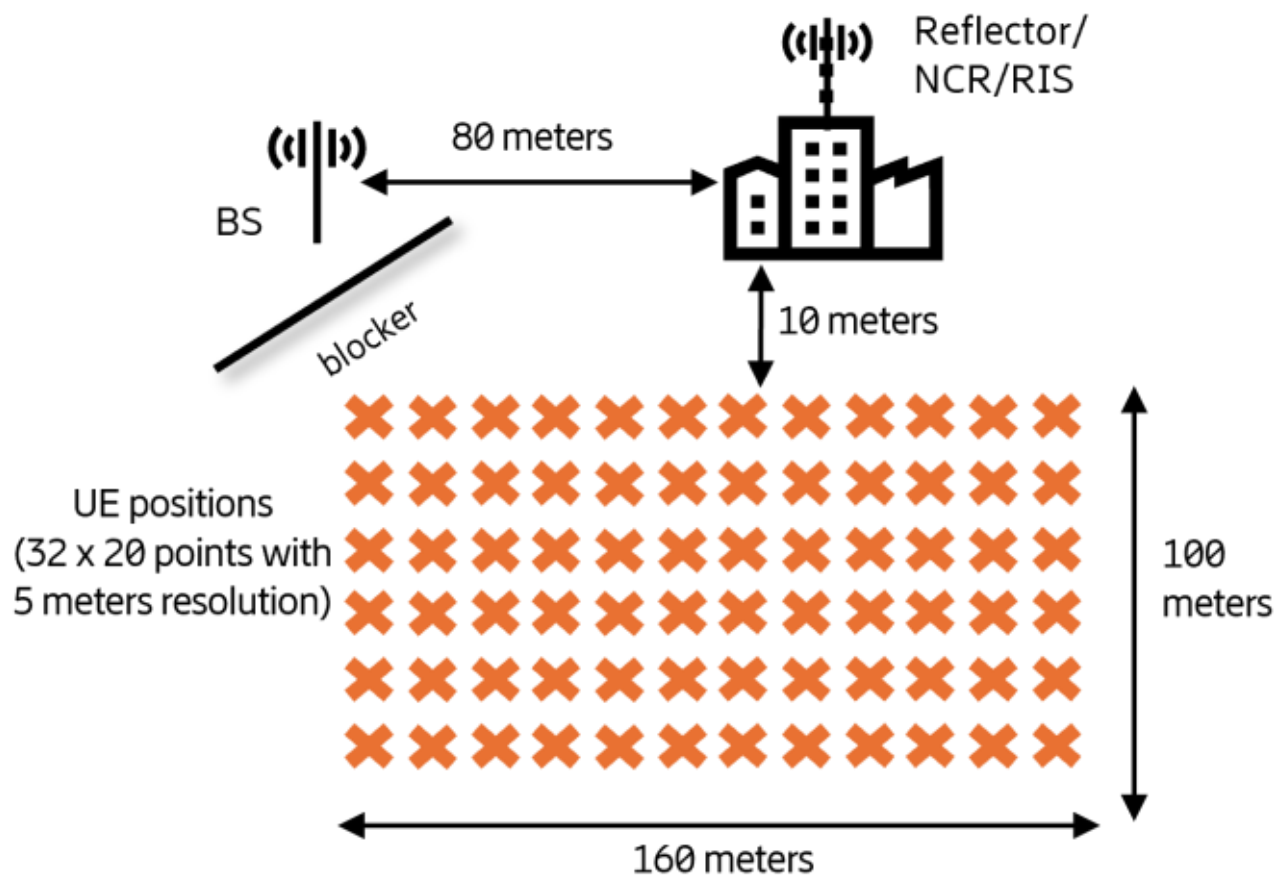

Figure 9. Simulation setup for blind-spot coverage scenario.

<u>**RIS**</u>: The phase shift matrix $\boldsymbol{\Phi}$**,** which is a diagonal matrix with unit-norm diagonal entries, is optimized. The channels between BS-RIS, RIS-UE, and BS-UE are denoted as $\mathbf{H}_{\mathrm{sr}}, \mathbf{h}_{\mathrm{rd}}, \mathbf{h}_{\mathrm{sd}}$, respectively. The beamformer for BS is denoted as $\mathbf{w}_{\mathrm{s}}$. We also consider external interference $\mathbf{z}_{\mathrm{ext}}$. The received signal at the UE can be written as

$$r_{\mathrm{d}} = \sqrt{P_{\mathrm{t}}}(\mathbf{h}_{\mathrm{rd}}\boldsymbol{\Phi}\mathbf{H}_{\mathrm{sr}} + \mathbf{h}_{\mathrm{sd}})\mathbf{w}_{\mathrm{s}}s + \mathbf{h}_{\mathrm{rd}}\boldsymbol{\Phi}\mathbf{z}_{\mathrm{ext}} + z_{\mathrm{d}}$$

that includes the signals coming from RIS and BS as well as the internal noise $z_{\mathrm{d}}$. Also, $P_{\mathrm{t}}$ is the total transmit power of the BS and $s$ is the information symbol sent to the UE. It is assumed that the external interference is a complex Gaussian noise vector with zero mean and covariance matrix $\sigma_{ext}^2\boldsymbol{R_{ext}}$.

[1] System-level simulations of NCR can be found in [40].

The internal noise is circularly symmetric complex Gaussian with variance $\sigma_d^2$. Considering desired signal, interference and noise terms, we can evaluate SINR and SE values as shown below

$$\mathrm{SINR}_{\mathrm{RIS}} = \frac{P_{\mathrm{t}}|(\mathbf{h}_{\mathrm{rd}}\boldsymbol{\Phi}\mathbf{H}_{\mathrm{sr}} + \mathbf{h}_{\mathrm{sd}})\mathbf{w}_{\mathrm{s}}|^2}{\sigma_{\mathrm{ext}}^2\left\|\mathbf{h}_{\mathrm{rd}}\boldsymbol{\Phi}\mathbf{R}_{\mathrm{ext}}{}^{1/2}\right\|^2 + \sigma_{\mathrm{d}}^2}$$

$$\mathrm{SE}_{\mathrm{RIS}} = \mathrm{E}[\log_2(1 + \mathrm{SINR}_{\mathrm{RIS}})]$$

For optimum SE, alternating optimization method is used where $\mathbf{w}_s$ and $\boldsymbol{\Phi}$ are iteratively optimized where

$$\widetilde{\mathbf{w}}_{\mathrm{s}} = (\mathbf{h}_{\mathrm{rd}}\boldsymbol{\Phi}\mathbf{H}_{\mathrm{sr}} + \mathbf{h}_{\mathrm{sd}})^H$$

$$\mathbf{w}_{\mathrm{s}} = \frac{\widetilde{\mathbf{w}}_{\mathrm{s}}}{\|\widetilde{\mathbf{w}}_{\mathrm{s}}\|}$$

At each step, one of them is optimized while keeping the other fixed. After the convergence, we quantize the phase shift values of $\boldsymbol{\Phi}$ and re-evaluate $\boldsymbol{w}_s$.

**Metal reflector**: We consider a reflector which does not change the amplitude of the reflected signal but randomly changes its phase. The received signal by the UE can be written as

$$r_{\mathrm{d}} = \sqrt{P_{\mathrm{t}}}(h_{\mathrm{rd}}e^{j\gamma}\mathbf{h}_{\mathrm{sr}} + \mathbf{h}_{\mathrm{sd}})\mathbf{w}_{\mathrm{s}}s + h_{\mathrm{rd}}e^{j\gamma}z_{\mathrm{ext}} + z_{\mathrm{d}}$$

where $\gamma$ is the random phase change introduced by the reflector which is generated as $\gamma \sim \mathrm{unif}(0,2\pi)$, and $z_{\mathrm{ext}} \sim CN(0, \sigma_{\mathrm{ext}}^2)$. The SINR and SE can be expressed as

$$\mathrm{SINR}_{\mathrm{reflector}} = \frac{P_{\mathrm{t}}\left|(h_{\mathrm{rd}}e^{j\gamma}\mathbf{h}_{\mathrm{sr}} + \mathbf{h}_{\mathrm{sd}})\mathbf{w}_{\mathrm{s}}\right|^2}{\sigma_{\mathrm{ext}}^2\|h_{\mathrm{rd}}\|^2 + \sigma_{\mathrm{d}}^2}$$

$$\mathrm{SE}_{\mathrm{reflector}} = \mathrm{E}[\log_2(1 + \mathrm{SINR}_{\mathrm{reflector}})]$$

and the BS beamformer will be optimized as

$$\widetilde{\mathbf{w}}_{\mathrm{s}} = (h_{\mathrm{rd}}e^{j\gamma}\mathbf{h}_{\mathrm{sr}} + \mathbf{h}_{\mathrm{sd}})^H$$

$$\mathbf{w}_{\mathrm{s}} = \frac{\widetilde{\mathbf{w}}_{\mathrm{s}}}{\|\widetilde{\mathbf{w}}_{\mathrm{s}}\|}$$

**NCR**: For full duplex operation, simultaneous transmission and reception is considered at the relay side, causing some self-interference (SI). The received signal equation at the NCR is expressed as

$$y_r = \mathbf{v}_1^H\left(\sqrt{P_{\mathrm{t}}}\mathbf{H}_{\mathrm{sr}}\mathbf{w}_{\mathrm{s}}s + \mathbf{z}_{\mathrm{si}} + \mathbf{z}_{\mathrm{ext}} + \mathbf{z}_{\mathrm{r}}\right)$$

where $\boldsymbol{v}_1$ is the receive beamformer of the relay, $\boldsymbol{z}_r$ is the internal noise vector at the relay, $\boldsymbol{z}_{si}$ is the SI term due to full-duplex operation. The received signal at the UE is written as

$$y_{\mathrm{d}} = \sqrt{P_{\mathrm{t}}}\mathbf{h}_{\mathrm{sd}}\mathbf{w}_{\mathrm{s}}s + c\mathbf{h}_{\mathrm{rd}}\mathbf{v}_2 y_r + z_{\mathrm{d}}$$

$$y_{\mathrm{d}} = \left(\sqrt{P_{\mathrm{t}}}\mathbf{h}_{\mathrm{sd}}\mathbf{w}_{\mathrm{s}} + c\mathbf{h}_{\mathrm{rd}}\mathbf{v}_2\mathbf{v}_1^H\sqrt{P_{\mathrm{t}}}\mathbf{H}_{\mathrm{sr}}\mathbf{w}_{\mathrm{s}}\right)s + c\mathbf{h}_{\mathrm{rd}}\mathbf{v}_2\mathbf{v}_1^H(\mathbf{z}_{\mathrm{si}} + \mathbf{z}_{\mathrm{ext}} + \mathbf{z}_{\mathrm{r}}) + z_{\mathrm{d}}$$

where $\boldsymbol{v}_2$ is the transmit beamformer of the NCR, the constant $c$ is the amplification factor of the NCR, $\boldsymbol{z}_d$ is the noise term at the UE. It is assumed that the self-interference is distributed as $z_{si} \sim CN(0, \sigma_{\mathrm{si}}^2\mathbf{I})$, here the SI power is $\sigma_{si}^2$ and this term includes path-loss between Tx and Rx of the NCR. We assume that LOS part of the SI is mitigated. Using the inequality

$$|\mathbf{h}_{\mathrm{sd}}\mathbf{w}_{\mathrm{s}} + c\mathbf{h}_{\mathrm{rd}}\mathbf{v}_2\mathbf{v}_1^H\mathbf{H}_{\mathrm{sr}}\mathbf{w}_{\mathrm{s}}| \le |\mathbf{h}_{\mathrm{sd}}\mathbf{w}_{\mathrm{s}}| + |c||\mathbf{h}_{\mathrm{rd}}\mathbf{v}_2\mathbf{v}_1^H\mathbf{H}_{\mathrm{sr}}\mathbf{w}_{\mathrm{s}}|$$

which is attained for a suitable angle of the amplification factor $c$, we can write the SINR and SE terms as

$$\mathrm{SINR}_{\mathrm{NCR}} = \frac{P_{\mathrm{t}}|(|\mathbf{h}_{\mathrm{sd}}\mathbf{w}_{\mathrm{s}}| + |c||\mathbf{h}_{\mathrm{rd}}\mathbf{v}_2\mathbf{v}_1^H\mathbf{H}_{\mathrm{sr}}\mathbf{w}_{\mathrm{s}}|)|^2}{\|c\mathbf{h}_{\mathrm{rd}}\mathbf{v}_2\mathbf{v}_1^H\|^2(\sigma_{\mathrm{si}}^2 + \sigma_{\mathrm{r}}^2) + \sigma_{\mathrm{ext}}^2\left\|c\mathbf{h}_{\mathrm{rd}}\mathbf{v}_2\mathbf{v}_1^H\mathbf{R}_{\mathrm{ext}}^{1/2}\right\|^2 + \sigma_{\mathrm{d}}^2}$$

$$\mathrm{SE}_{\mathrm{NCR}} = \mathrm{E}[\log_2(1 + \mathrm{SINR}_{\mathrm{NCR}})]$$

The beamformers and $c$ are optimized as

$$\mathbf{w}_{\mathrm{s}} = \mathrm{prsv}(\mathbf{H}_{\mathrm{sr}}), \qquad \mathbf{v}_1 = \mathbf{H}_{\mathrm{sr}}\mathbf{w}_{\mathrm{s}} \quad \mathbf{v}_2 = \frac{\mathbf{h}_{\mathrm{rd}}^H}{\|\mathbf{h}_{\mathrm{rd}}\|}$$

where $\mathrm{prsv}(\cdot)$ finds the principal right singular vector. The maximum $|c|^2$ will be found such that

$$|c|^2\|\mathbf{v}_2\|^2(P_{\mathrm{t}}|\mathbf{v}_1^H\mathbf{H}_{\mathrm{sr}}\mathbf{w}_{\mathrm{s}}|^2 + \sigma_{\mathrm{si}}^2\mathbf{v}_1^H\mathbf{v}_1 + \sigma_{\mathrm{ext}}^2\mathbf{v}_1^H\mathbf{R}_{\mathrm{ext}}\mathbf{v}_1 + \sigma_{\mathrm{r}}^2\mathbf{v}_1^H\mathbf{v}_1) \le P_{\mathrm{t},2}$$

$$|c|^2 \le G_{\max}$$

where $G_{max}$ is the maximum amplification factor of the NCR.

Coverage heat maps in terms of spectral efficiency (SE) as a function of the UE position are given in Fig. 10. In these maps, the BS is located at the origin (0, 0) and the reflector/relay/RIS is deployed at the relative location (80m, 0m) according to the simulation setup with the 15 GHz operating frequency. Firstly, we observe that simple reflector case shows that reflection without any amplification/gain cannot enhance the SE values. For the case when an intelligent surface is deployed but with a fixed pattern, so-called fixed RIS, i.e., non-reconfigurable as RIS phase shift matrix optimized as if the UE is at the center of the region, good performance is observed only when the angle of departure from RIS to UE is close to the angle obtained when the UE is at the center; for other UE locations, we do not observe any gain by fixed RIS, compared to simple reflector case. Further, when we compare RIS without any fixed pattern constraint with NCR, we observe that RIS can provide similar or slightly better performance than NCR when UE is close to the RIS/relay, however NCR not only outperforms RIS when the UE is not close but also provides a uniform performance showing that the relay performance is interference-limited. One should note that, given that self-interference suppression capability increases, relay performance can be improved. It is concluded that the performance of RIS depends on RIS-to-UE distance and hence RIS performance is noise-limited.

Table III.
Simulation parameters related to Figs. 9, 10 and 11.

| **Parameter** | **Value** |
|---|---|
| Frequency | 15, 28 GHz |
| Bandwidth | 100, 400 MHz |
| BS, RIS/relay, UE heights | 20, 2, 1.5 m |
| Small-scale fading model | LoS: Rician |
| Noise Figure | 10 dB |
| Number of BS antennas | 64 |
| Number of RIS elements | 256 |
| Number of relay antennas | 16 |
| BS transmit power | 40 dBm |
| Maximum relay transmit power | 40 dBm |
| Ratio of external interference power and noise power at relay | 10 dB |
| Self-interference suppression | 90 dB |

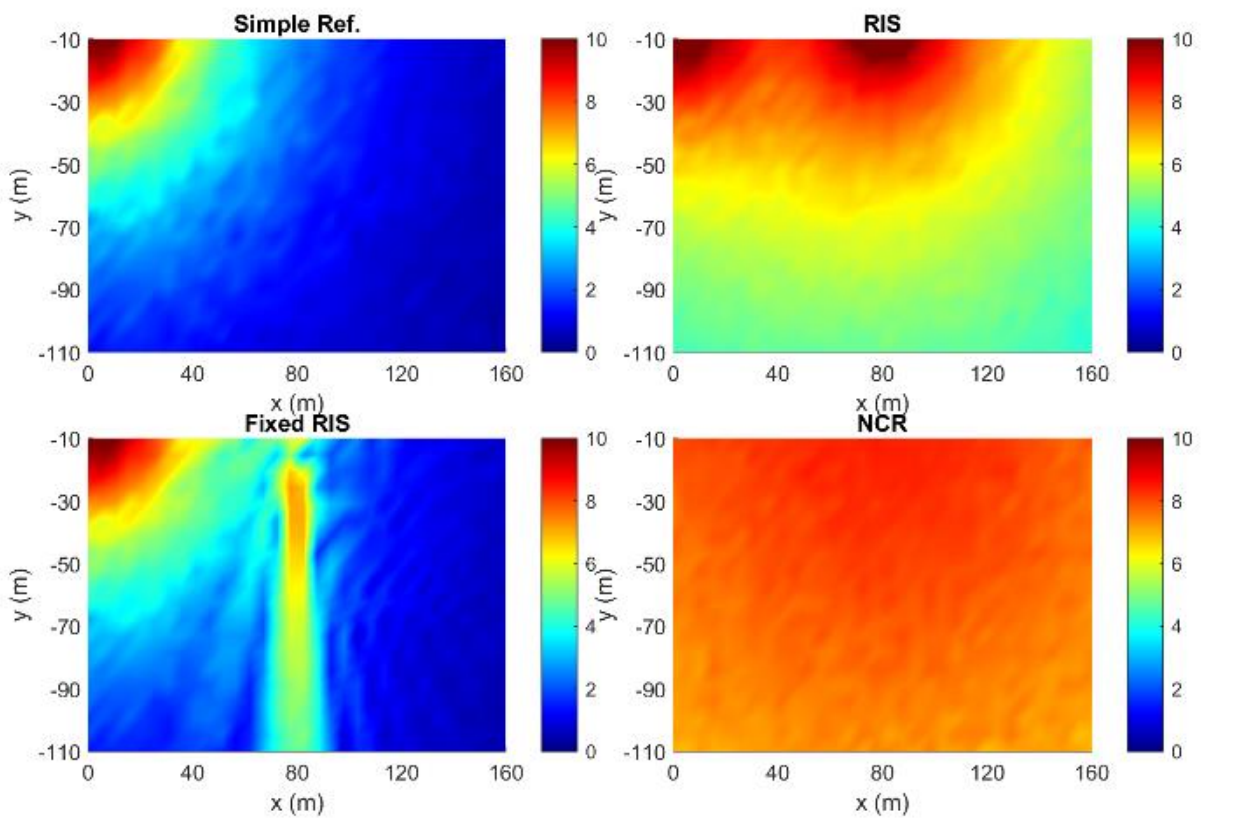

**Figure 10.** **SE coverage heat maps for simple reflector, RIS, fixed RIS and NCR for a blind-spot coverage scenario at 15 GHz.**

Figure 11 shows the corresponding cumulative distribution function (CDF) curves of SE values obtained for simple reflectors, fixed RIS, RIS and NCR operating at both 15 GHz and 28 GHz. The simulations are based on the parameter setting of Table 3. It is observed that NCR outperforms RIS at both frequencies, where the simple reflector and fixed RIS show the worst performance. As the operating frequency increases, the performance decreases, e.g., mean SE for NCR and RIS decreases by 5.3 bps/Hz and 4.8 bps/Hz, respectively, while significant performance degradation is seen for lower percentiles.

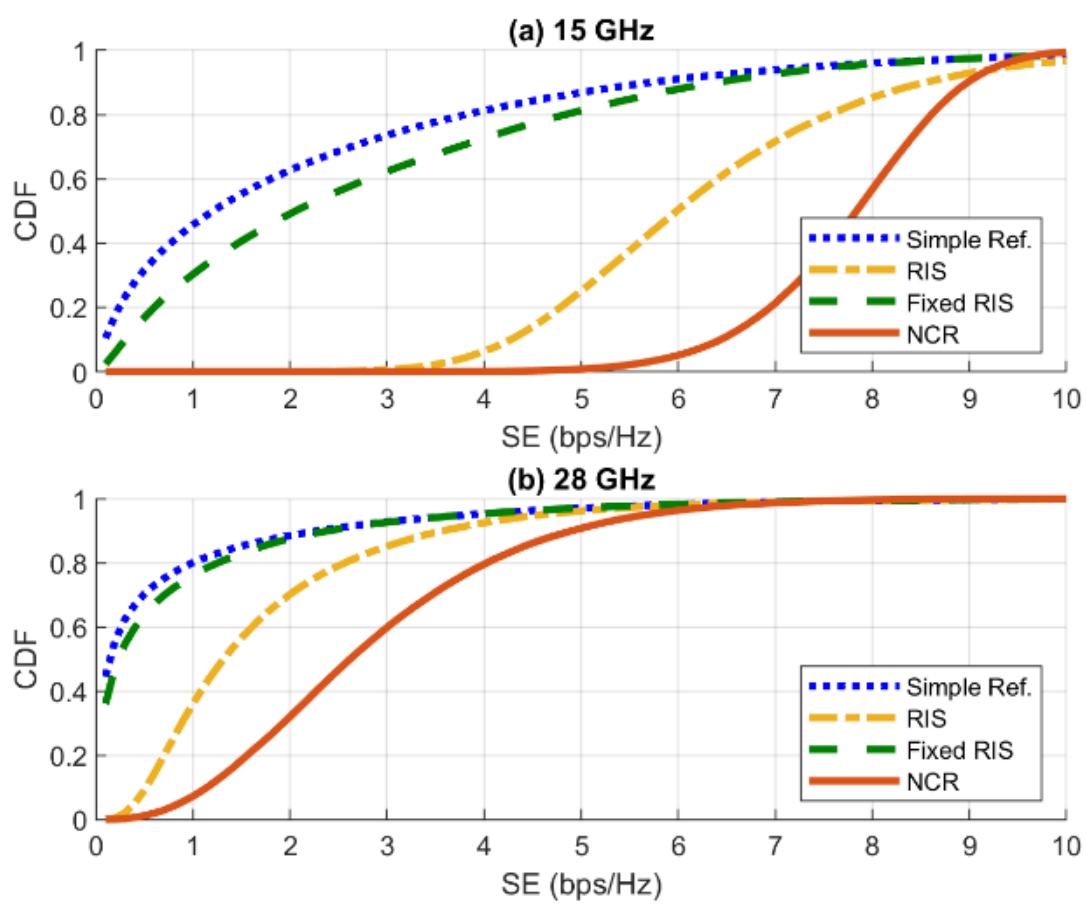

**Figure 11.** **SE CDFs for simple reflector, RIS, fixed RIS and NCR for a blind-spot coverage scenario at (a) 15 GHz and (b) 28 GHz.**

## VI. CONCLUSION

In this paper, we studied the issues and challenges of RIS which may affect the feasibility/usefulness of RIS in cellular networks. Particularly, we presented the intuitive understandings which have resulted in not yet considering RIS in the 3GPP standardizations. Also, we presented various simulation results to highlight the pros and cons of RIS, particularly in comparison with alternative technologies. As explained, RIS is more of a 6G technology, and multiple practical issues need to be addressed before it can be used in large scale in cellular networks.

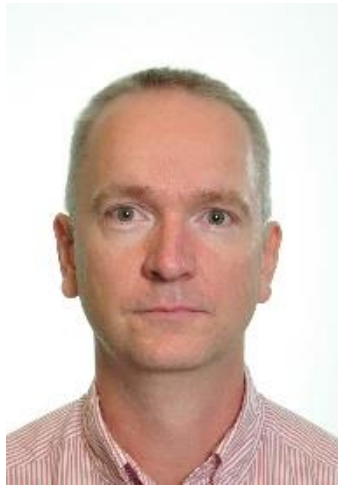

**MAGNUS ÅSTRÖM** received his Ph.D. degree in digital signal processing from Lund University in 2003. Following that, he has held positions within R&D at a number of companies working with different wireless technologies, e.g., Bluetooth, 802.11, GSM/GPRS, WCDMA/HSPA, LTE and NR. He is presently a research manager and 3GPP RAN1 delegate at Ericsson Research with contributions in machine-type communication, narrowband-IoT, integrated access and backhaul, network-controlled repeaters and presently subband full duplex and various 6G research topics. Magnus is the inventor of over 150 filed or granted patents.

**Philipp Karl Gentner** was born in Laupheim, Germany in January 1980. He received his Diploma in electrical engineering in 2006 at the University of Ulm. In 2013 he received his Dr. techn. degree from the Institute of Telecommunications in Vienna, Austria. He gained a lot of practical experience during his studies by working for Siemens Mobile Phones and later BenQ-Siemens. As a System Test Engineer for Ubidyne he was responsible for the System Test Laboratory and the definition and implementation of RF Test Cases for the first active antenna systems. Since 2014 he is working on antenna architectures for future communication systems at Ericsson Antenna Systems Germany. Currently he holds a senior researcher position, and his research interests are in the computer aided engineering (digital twinning) of antenna systems and new methodologies for antenna element design.

**Omer Haliloglu** received his Ph.D. in 2018 from Electrical and Electronics Engineering from Hacettepe University, Turkiye. He has been with Ericsson Research since 2018 and conducting research in the radio networks area with a focus on MIMO, multi-node connectivity, flexible deployment topologies and artificial intelligence.

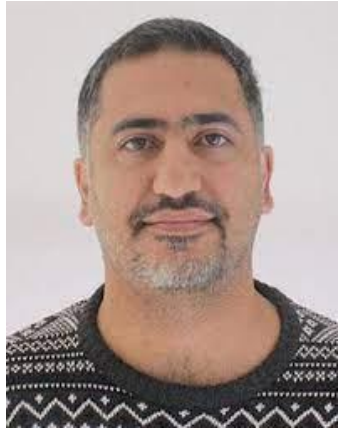

**BEHROOZ MAKKI** (Senior Member, IEEE) received the Ph.D. degree in communication engineering from the Chalmers University of Technology, Gothenburg, Sweden. He was a Postdoctoral Researcher with the Chalmers University of Technology from 2013 to 2017. He is currently working as a Senior Researcher with Ericsson Research, Gothenburg. He has coauthored more than 70 journal articles, more than 50 conference papers, and more than 130 patent applications. His current research interests include different aspects of 5G/6G wireless networks. Behrooz serves as an Editor in IEEE Wireless communication letters, IEEE Transactions on Wireless communications, and IEEE Transactions on Communications as well as Senior Editor in IEEE Communications Letters

**Ola Tageman** received the M.Sc. degree in engineering physics and the Ph.D. degree in theoretical physics from the Chalmers University of Technology, Gothenburg, Sweden, in 1993 and 1998, respectively. Since 1999, he has been with Ericsson, Gothenburg, in different roles, and currently holds an Expert position on integrated millimeter wave subsystems within Ericsson Research. His research interests include antenna systems integration, electromagnetic modeling, design automation, millimeter wave front ends, array antennas, filters, packaging, interconnects, assembly, tunable RF technologies, sub-THz commmunications and full-duplex transmission.